\documentstyle[12pt]{article}

\newcommand{\sect}[1]{\setcounter{equation}{0}\section{#1}}

\newcommand{\bea}{\begin{eqnarray}}
\newcommand{\eea}{\end{eqnarray}}
\newcommand{\be}{\begin{equation}}
\newcommand{\ee}{\end{equation}}
\newcommand{\vs}[1]{\vspace{#1 mm}}

\renewcommand{\a}{\alpha}
\renewcommand{\b}{\beta}

\renewcommand{\d}{\delta}
\newcommand{\e}{\epsilon}
\newcommand{\dsl}{\pa \kern-0.5em /}

\newcommand{\pa}{\partial}

\newcommand{\nn}{\nonumber\\}

\begin{document}
\topmargin 0pt
\oddsidemargin 0mm

\begin{flushright}
USTC-ICTS-04-19\\
MCTP-04-51\\
hep-th/0408242\\
\end{flushright}

\vs{2}
\begin{center}
{\Large \bf
Static, non-SUSY $p$-branes in diverse
dimensions}
\vs{10}

{\large J. X. Lu$^a$\footnote{E-mail: jxlu@ustc.edu.cn}
 and Shibaji Roy$^b$\footnote{E-mail: roy@theory.saha.ernet.in}}

 \vspace{5mm}

{\em
 $^a$ Interdisciplinary Center for Theoretical Study\\
 University of Science and Technology of China, Hefei, Anhui 230026,
P. R. China\\
and\\
Interdisciplinary Center of Theoretical Studies\\
Chinese Academy of Sciences, Beijing 100080, China\\
 and\\

Michigan Center for Theoretical Physics\\
Randall Laboratory, Department of Physics\\
University of Michigan, Ann Arbor, MI 48109-1120, USA\\

\vs{4}

 $^b$ Saha Institute of Nuclear Physics,
 1/AF Bidhannagar, Calcutta-700 064, India}
\end{center}

\vs{5}
\centerline{{\bf{Abstract}}}
\vs{5}
\begin{small}
We give explicit constructions of static, non-supersymmetric $p$-brane
(for $p \leq d-4$, where $d$ is the space-time dimensionality and including
$p=-1$ or D-instanton) solutions of type II supergravities in 
diverse dimensions. A subclass of these are the static
counterpart of the time dependent solutions obtained in [hep-th/0309202].
Depending on the forms of the non-extremality function $G(r)$ defined
in the text, we discuss various possible solutions and their region of
validity. We show 
how one class of these solutions interpolate between the $p$-brane--
anti $p$-brane 
 solutions
 and the usual BPS $p$-brane solutions in
$d=10$, while the other class, although have BPS limits, do not have 
such an interpretation. 
We point out how the time dependent solutions mentioned above can
be obtained by a Wick rotation of one class of these static solutions. We also
discuss another type of solutions which might seem non-supersymmetric, but
we show by a coordinate transformation that they are nothing but the near
horizon limits of the various BPS $p$-branes already known.
\end{small}
\newpage

\section{Introduction}

Type II supergravities admit static, supersymmetric space-time geometries
with isometries ISO($p,\,1$) $\times$ SO($d-p-1$) in $d$-dimensions known
as BPS $p$-branes \cite{hs,dl}. If one wants to construct an 
analogous time-dependent
geometries with isometries ISO($p+1$) $\times$ SO($d-p-2,\,1$), one finds
that there are no real Euclidean $p$-brane (or S-brane) solutions in this
case \cite{br}. However, if we do not insist on supersymmetries, 
then there exist
real solutions with metrics having the aforementioned isometries \cite{br}. 
So, it would be natural to ask whether there is an analogous
static, non-supersymmetric $p$-brane solutions with isometries 
ISO($p,\,1$) $\times$ SO($d-p-1$) in type II supergravities
in arbitrary $d$ space-time dimensions and we find the answer in the positive.
These solutions are not of the type of black $p$-branes \cite{hs} which 
are also
non-supersymmetric but have isometries R $\times$ ISO($p$) $\times$
SO($d-p-1$).
We give explicit constructions of these
solutions. Although a subclass of these solutions were
previously known \cite{zz,im,mo}, in a 
different form, we give explicit constructions to facilitate our
discussion on certain aspects of these solutions not considered before.

We construct the solutions by solving the equations of motion of type II
supergravities in $d$ space-time dimensions containing a graviton, a dilaton
and a $(d-p-3)$-form gauge field. It is well-known that when the 
supersymmetry condition is imposed the equations of motion lead to the
usual BPS $p$-brane solutions \cite{hs,dl,dkl}. However, in analogy 
with 
the time-dependent solutions \cite{br}
we relax the supersymmetry condition by introducing a non-extremality
function $G(r)$ (defined below) and find a real magnetically (electrically)
charged
$p$-brane (for $-1 \leq p \leq 6$) solutions which are characterized by 
three or less number of parameters. For the consistency of the equations 
of motion, we find that the non-extremality function can not be arbitrary
and should take some specific forms. Demanding the asymptotic flatness
of the metric we find that the non-extremality function $G(r)$ can be of
the forms $G_{\mp} (r) = 1 \mp \omega^{2(d-p-3)}/r^{2(d-p-3)}$, where
$\omega$ is a real integration constant. The upper sign leads to the
three or two parameter static, non-supersymmetric $p$-brane solutions, whereas
the lower sign leads to only two parameter solutions. Usually these 
solutions have
singularities and we discuss the region of validity for these solutions. Since
$p=-1$ or the case of D-instanton is quite different from the rest of the
$p$-brane solutions we discuss it separately. Then we clarify the relations
between the three parameter solutions and those obtained in \cite{zz}. Next
we show that when the non-extremality function has the upper sign, the
three parameter solutions nicely interpolate between the chargeless 
$p$-brane--anti $p$-brane system 
and the usual BPS $p$-branes by scaling the parameters of the solutions in two
distinct ways for $0\leq p \leq 6$ and in a unique way for $p=-1$.
This does not happen for the solutions with the non-extremality
function having the lower sign. However, we find that  even in this case it is 
possible to obtain the BPS solutions by appropriately scaling the parameters
for $0\leq p \leq 6$ but not for $p=-1$ or D-instanton. These solutions are
not of the type of the usual BPS $p$-branes. It should be emphasized that here
we consider these solutions
as just the classical supergravity solutions and will not
try to give any microscopic string interpretation. One such possible
interpretation was given in \cite{bmo} by considering these solutions
(actually a subclass of these solutions (the three parameter solutions)
with the non-extremality function
having the upper sign)
in $d=10$ as the coincident D$p$--${\bar {\rm D}}p$ branes and the
three parameters
of these solutions were argued to be related to the numbers $N$ of D$p$-branes,
${\bar N}$ of ${\bar {\rm D}}p$-branes and the tachyon vev ($T$) of the
brane-antibrane system. Recently we proposed \cite{luroy} an exact 
relationships of the
parameters of these solutions to the physically relevant parameters $N$,
${\bar N}$ and $T$ and have shown how these relations are consistent with
the right picture of tachyon condensation \cite{astwo} on the 
brane--anti brane system.
Time-dependent solutions can sometimes be obtained from the static solutions
by applying Wick rotation. While it is known that
the Wick rotations of the static BPS $p$-brane solutions do not
lead to real time-dependent solutions, we show how a subclass of 
these non-supersymmetric
solutions lead to the real time-dependent solutions obtained in \cite{br}.
Finally, we also discuss another type of solutions where we do not
demand the asymptotic flatness of the metric and take the non-extremality
function to be of the form $G(r) = \omega^{2(d-p-3)}/r^{2(d-p-3)}$. 
Although these solutions apparently seem to
be non-supersymmetric, but actually in $d=10$ they can be shown
by a coordinate transformation to be the near-horizon limits of various BPS
$p$-branes we know \cite{malda,imsy}. 

This paper is organized as follows. In section 2, we describe the construction
of static, non-supersymmetric $p$-brane solutions in $d$-dimensional
supergravities. Various aspects of these solutions are discussed in section 3.
In section 4, we obtain another class of solutions which are shown to be
the near horizon limits of various BPS $p$-branes by a coordinate
transformation. Our conclusion is presented in section 5.

\section{General static, non-SUSY $p$-branes}

In this section we describe the construction of static, non-supersymmetric
$p$-branes by solving the equations of motion of type II supergravities in
$d$ dimensions. The $d$-dimensional supergravity action containing a metric,
a dilaton and a $(q-1)=(d-p-3)$-form gauge field with dilaton coupling $a$
has the form,
\be
S = \int d^dx \sqrt{-g} \left[R - \frac{1}{2} \partial_\mu \phi \partial^\mu
\phi - \frac{1}{2\cdot q!} e^{a\phi} F_{[q]}^2\right]
\ee
The above action is quite general and consists of the bosonic sector of
(dimensionally reduced) string/M theories.
The equations of motion following from (2.1) are,
\bea
R_{\mu\nu} - \frac{1}{2} \partial_\mu \phi\partial_\nu \phi -
\frac{e^{a\phi}}
{2(q-1)!} \left[F_{\mu\alpha_2\ldots \alpha_q}F_\nu^{\,\,\,\alpha_2
\ldots \alpha_q}
- \frac{q-1}{q(d-2)}F^2_{[q]}g_{\mu\nu}\right] &=& 0\\
\partial_\mu\left(\sqrt{-g}e^{a\phi} F^{\mu\alpha_2\ldots
\alpha_q}\right) &=& 0\\
\frac{1}{\sqrt{-g}}\partial_\mu\left(\sqrt{-g}\partial^\mu \phi\right)
- \frac{a}{2 \cdot q!} e^{a\phi} F_{[q]}^2 &=& 0
\eea
We will solve the equations of motion with the following ansatz,
\bea
ds^2 &=& e^{2A(r)}\left(dr^2 + r^2 d\Omega_{d-p-2}^2\right) + e^{2B(r)} \left(
-dt^2 + dx_1^2 + \cdots + dx_{p}^2\right)\\
F_{[q]} &=& b\,\, {\rm Vol}(\Omega_{d-p-2})
\eea
In the above $r = (x_{p+1}^2 + \cdots + x_{d-1}^2)^{1/2}$, $d\Omega_{d-p-2}^2$
is the line element of a unit $(d-p-2)$-dimensional sphere,
Vol($\Omega_{d-p-2}$) is its volume-form and $b$ is the magnetic charge
parameter. The space-time in (2.5) has the isometry SO($d-p-1$) $\times$
ISO($p,\,1$) and therefore the above represent a magnetically charged
$p$-brane in $d$ dimensions. It is well known that the solution is
supersymmetric saturating the BPS bound if the function $A(r)$ and $B(r)$
satisfy
\be
(p+1) B(r) + (q-1) A(r) = 0
\ee
Actually the condition of the preservation of some fraction of supersymmetries
gets translated to the above condition on the metric and was shown in 
\cite{dkl},
with $p+1 = d$ and $q-1 = \tilde d$ in their notation. It is well-known that
the solutions of the equations of motion with (2.5) -- (2.7) lead to the
usual BPS $p$-branes \cite{hs,dl,dkl}. We will relax the 
condition (2.7)
by
\be
(p+1) B(r) + (q-1) A(r) = \ln G(r)
\ee
As long as $G(r) \neq 1$, we expect the solution to break all the space-time
supersymmetries. However, we will give an example in the last section where
$G(r) \neq 1$, still one can make a coordinate transformation and modify
$A(r)$ accordingly such that the relation (2.8) reduces to the form (2.7)
and the supersymmetry will be restored. This does not happen for the solutions
considered in this section.

The non-vanishing Ricci tensor components can be obtained from (2.5) as,
\bea
R_{rr} &=& (p+1) \left[B'' + B'^2 - A'B'\right] +
q\left[A'' + \frac{A'}{r}\right]\\
R_{xx} &=& -R_{tt}\,\,\,=\,\,\, e^{2B-2A}\left[B'' + (q-1) A'B' + (p+1) B'^2
+ q \frac{B'}{r}\right]\\
R_{ab} &=& r^2\left[A'' + (q-1) A'^2 + (2q-1) \frac{A'}{r}
+ (p+1) B' (A' + \frac{1}{r})\right]\bar{g}_{ab}
\eea
where $a,\,b$ are the indices for the transverse spherical (angular)
coordinates and ${\bar g}_{ab}$ is the metric for the unit
$q=(d-p-2)$-dimensional sphere. $x$ denotes the indices for the longitudinal
directions. Also `prime' here denotes the derivative with respect to $r$.
With the ansatz (2.6), eq.(2.3) is automatically satisfied. We rewrite the
other equations of motion (2.2) and (2.4) using (2.9) -- (2.11) and (2.8) as
follows,
\bea
A'' + \frac{G''}{G} - \frac{G'^2}{G^2} + \frac{1}{p+1}\left(\frac{G'}{G}
- (q-1)A'\right)^2 + (q-1) A'^2 - \frac{G'}{G} A' + \frac{q}{r} A' & & \nn
+ \frac{1}{2} {\phi'}^2
- \frac{b^2(q-1)}{2(d-2)} \frac{e^{2(p+1)B + a\phi}}{G^2 r^{2q}} &=& 0
\qquad\qquad\\
B'' + \frac{q}{r}B' + \frac{G'}{G} B' - \frac{b^2(q-1)}{2(d-2)}
\frac{e^{2(p+1)B+a\phi}} {G^2 r^{2q}} &=& 0\qquad\qquad \\
A'' + \frac{q}{r}A' + \frac{G'}{G} (A'+\frac{1}{r}) +
\frac{b^2(p+1)}{2(d-2)} \frac{e^{2(p+1)B+a\phi}} {G^2 r^{2q}} &=& 0
\qquad\qquad\\
{\phi''} + \frac{q}{r}{\phi'} + \frac{G'}{G}{\phi'} -
\frac{a b^2}{2}\frac{e^{2(p+1)B+a\phi}}{G^2 r^{2q}} &=& 0\qquad\qquad
\eea
Expressing eq.(2.14) in terms of $B(r)$ using (2.8) and substituting the
equations of motion for $B(r)$ (eq.(2.13)), we get an equation involving
the function $G(r)$ only as,
\be
G'' + \frac{2q-1}{r} G' = 0
\ee
Assuming $G(r)$ to go to unity asymptotically we find two solutions of the
above equation as,
\be
G_-(r) = 1 - \frac{\omega^{2(q-1)}}{r^{2(q-1)}}, \qquad
G_+(r) = 1 + \frac{\tilde{\omega}^{2(q-1)}}{r^{2(q-1)}}
\ee
where both $\omega$ and $\tilde{\omega}$ are real. We factorize $G_-(r)$ 
and $G_+(r)$ as follows,
\bea
G_-(r) &=& 1 - \frac{\omega^{2(q-1)}}{r^{2(q-1)}} = \left(1+\frac{\omega^{q-1}}
{r^{q-1}}\right)\left(1-\frac{\omega^{q-1}}{r^{q-1}}\right) =
H_1 (r) {\tilde H}_1 (r)\nn
G_+(r) &=& 1 + \frac{\tilde{\omega}^{2(q-1)}}{r^{2(q-1)}} = \left(1+\frac{i
\tilde{\omega}^{q-1}}
{r^{q-1}}\right)\left(1-\frac{i\tilde{\omega}^{q-1}}{r^{q-1}}\right) =
H_2 (r) {\tilde H}_2 (r)
\eea
where $H_1(r) = 1 + \omega^{q-1}/r^{q-1}$, ${\tilde H}_1(r) = 
1 - \omega^{q-1}/
r^{q-1}$, 
$H_2(r) = 1 + i\tilde{\omega}^{q-1}/r^{q-1}$ and ${\tilde H}_2(r) = 
1 - i\tilde{\omega}^{q-1}/
r^{q-1}$. We first solve the equations of motion with $G_-(r)$ having the form
given in eq.(2.18). First using (2.13) and (2.15) we find
\be
\left({\phi} - \frac{a(d-2)}{q-1}B\right)''
+ \frac{q}{r}\left({\phi} - \frac{a(d-2)}{q-1}B\right)'
+ \frac{G_-'}{G_-}\left({\phi} - \frac{a(d-2)}{q-1}B\right)' = 0
\ee
The solution to this equation takes the form,
\be
\phi = \frac{a(d-2)}{q-1} B + \delta \ln \frac{H_1}{\tilde H_1}
\ee
where $\delta$ is an arbitrary real constant. Now using (2.20)
we express
\be
e^{2(p+1)B+a\phi} = \left(\frac{H_1}{\tilde H_1}\right)^{a\delta} e^{B\chi}
\ee
where $\chi = 2(p+1) + a^2(d-2)/(q-1)$ and the equation for the function
$B(r)$ in (2.13) takes the form,
\be
B'' + \frac{q}{r}B' + \frac{G_-'}{G_-} B' - \frac{b^2(q-1)}{2(d-2)}
\frac{e^{B\chi} H_1^{a\delta - 2}} {r^{2q} {\tilde H}_1^{a\delta + 2}} = 0
\ee
Following the arguments given in ref.\cite{br}, we make the following ansatz
for $B(r)$,
\be
e^B = \left[\cosh^2\theta \left(\frac{H_1}{\tilde {H}_1}\right)^\alpha -
\sinh^2\theta
\left(\frac{\tilde {H}_1}{H_1}\right)^\beta\right]^\gamma = F_1^\gamma
\ee
where $F_1 = \left[\cosh^2\theta \left(\frac{H_1}{\tilde {H}_1}\right)^\alpha
- \sinh^2\theta
\left(\frac{\tilde {H}_1}{H_1}\right)^\beta\right]$, with $\alpha$, $\beta$,
$\theta$ being some parameters and $\gamma$ is another parameter
which will be determined shortly. Note that the ansatz (2.23) differs
from the similar ansatz (3.18) of ref.\cite{br} for the time-dependent case.
The reason we have hyperbolic function here instead of trigonometric function
is that there is a sign difference in the last term of (2.22) from the
corresponding equation in the time-dependent case. This will prove to be
crucial to recover the BPS $p$-brane solution in this case
(this was not possible for the time-dependent case as there is no BPS S-brane
solution in type II string theories) in the next section.
Substituting (2.23) in (2.22) we obtain,
\be
\left[\gamma (q-1) \omega^{2(q-1)} (\alpha+\beta)^2 \sinh^2 2\theta
\right]\frac{H_1^{\alpha-\beta} {\tilde {H}_1}^{\beta-\alpha}}{F_1^2}
+\frac{b^2}{2(d-2)} H_1^{a\delta} {\tilde {H}_1}^{-a\delta}F_1^{\gamma\chi} = 0
\ee
We thus obtain from here
\bea
\gamma \chi &=& -2, \qquad\qquad \alpha-\beta \,\,=\,\, a\delta\nn
b &=& \sqrt{\frac{4 (q-1) (d-2)}{\chi}} (\alpha+\beta) \omega^{q-1}
\sinh 2\theta
\eea
We have taken $\alpha+\beta$, $b$ and $\theta$ $\geq 0$ without any loss
of generality.
From (2.25) we note that the parameter $\gamma$ gets fixed and among $\alpha$,
$\beta$, $\delta$ only two are independent. However the consistency of the
equations of motion (2.12) gives a relation between the parameter $\alpha$ and
$\delta$ as,
\be
\frac{1}{2} \delta^2 + \frac{2\alpha(\alpha-a\delta)(d-2)}{\chi(q-1)}
= \frac{q}{q-1}
\ee
From (2.26) we can determine both $\alpha$ and $\beta$ in terms of $\delta$
as,
\bea
\alpha &=& \sqrt{\frac{\chi q
}{2(d-2)} - \frac{\delta^2}{4}\left(\frac{\chi(q-1)}{d-2} - a^2\right)}
+ \frac{a\delta}{2}\nn
\beta &=& \sqrt{\frac{\chi q
}{2(d-2)} - \frac{\delta^2}{4}\left(\frac{\chi(q-1)}{d-2} - a^2\right)}
- \frac{a\delta}{2}
\eea
Note in the above that even though $\delta$ is real, the parameters $\a$
and $\b$ are not necessarily real. In fact depending on the value of $\delta$
we have two cases,
\bea
(i) && |\delta|\,\, \leq \,\, \sqrt{\frac{\chi q}{(q-1)(p+1)}}, \quad
{\rm then}\,\,\, \a,\,\,\b\,\,\,{\rm are\,\, both\,\, real}\nn
(ii) && |\delta|\,\, > \,\, \sqrt{\frac{\chi q}{(q-1)(p+1)}}, \quad
{\rm then}\,\,\, \a,\,\,\b\,\,\,{\rm are\,\, both\,\, complex}
\eea
For case $(i)$ $|\d|$ is bounded, on the other hand for case $(ii)$
$|\d|$ can be arbitrarily large.
We thus obtain from (2.25) and (2.8)
\bea
e^{2B} &=& F_1^{-\frac{4}{\chi}}\nn
e^{2A} &=& (H_1 {\tilde {H}_1})^{\frac{2}{q-1}} F_1^{\frac{4(p+1)}{(q-1)\chi}}
\eea
and the complete non-supersymmetric, static, magnetically charged $p$-brane
solutions as,
\bea
ds^2 &=& F_1^{\frac{4(p+1)}{(q-1)\chi}} (H_1{\tilde {H}_1})^{\frac{2}{q-1}}
\left(dr^2 + r^2
d\Omega_{d-p-2}^2\right) + F_1^{-\frac{4}{\chi}}\left(-dt^2 + dx_1^2 + \cdots +
dx_{p}^2\right)\nn
e^{2\phi} &=& F_1^{-\frac{4a(d-2)}{(q-1)\chi}}
\left(\frac{H_1}{\tilde {H}_1}\right)^{2\delta}\nn
F_{[q]} &=& b\,\,{\rm Vol}(\Omega_{d-p-2})
\eea
There are three independent paramaters $\delta$, $\omega$ and $\theta$
characterizing the solutions (for case $(i)$ above, see the discussion below). 
These solutions have some similarities with
the BPS $p$-brane solutions in $d$ dimensions. In fact, if $H_1$, 
${\tilde {H}_1}$
$\to$ 1 and $F_1 \to $ the usual harmonic function, then these solutions indeed
reduce to the magnetically charged BPS $p$-brane solutions. We will come back
to it in more detail in the next section.

The solutions (2.30) represent the magnetically charged $p$-branes. The
corresponding electrically charged branes can be obtained from these
solutions by using the transformation $g_{\mu\nu} \to g_{\mu\nu}$,
$\phi \to -\phi$ and $F \to e^{-a\phi} \ast F$, where $\ast$ denotes the
Hodge dual. So, for the electrically charged solutions the field strength 
can be
calculated from above as,
\be
F_{[p+2]} = e^{a\phi} \ast F_{[q]}
\ee
where the dilaton is as given in eq.(2.30). The $(p+1)$-form gauge field
can be calculated from (2.31) as,
\be
A_{[p+1]} = \sqrt{\frac{4(d-2)}{\chi(q-1)}} \sinh\theta \cosh\theta
\left(\frac{C_1}{F_1}\right) dt\wedge \cdots \wedge dx_p
\ee
where,
\be
C_1 = \left(\frac{H_1}{\tilde {H}_1}\right)^\alpha - \left(\frac{\tilde {H}_1}
{H_1}\right)^
\beta
\ee
Note that as long as $|\d|$ is bounded as $(i)$ in (2.28), $\a$ and $\b$
are real and so, $F_1$ given in (2.23) is also manifestly real and
positive. 
But for case
$(ii)$ in (2.28) $\a$ and $\b$ are both complex and so, $F_1$ is not manifestly
real. In this case let us write $\a=i c+a\d/2$ and 
$\b=i c-a\d/2$ where $ c =
\sqrt{
\frac{\delta^2}{4}\left(\frac{\chi(q-1)}{d-2} - a^2\right)-
\frac{\chi q}{2(d-2)}}$ = positive\footnote{One can show this, for
example, by noting
that $a^2 = 4 - \frac{2 (p + 1)(q - 1)}{d - 2}$ for supergravities with
maximal susy in diverse dimensions\cite{dl}}. Then we find from eq.(2.25) that since 
$(\a+\b)=2i c$ purely imaginary $b$ will be real positive 
only for $\theta
= - i\tilde{\theta}$. It can be easily checked that $F_1$ in this case will
be real only for $\tilde{\theta} = \pi/4$ and takes the form,
$F_1 = exp\{a\d \tanh^{-1}(\omega^{q-1}/r^{q-1})\}
\cos\left(2 c\tanh^{-1}
(\omega^{q-1}/r^{q-1})\right)$.
With this $F_1$ the solution has the same form as in eq.(2.30). The gauge
field in this case takes the form
$A_{[p+1]} = \sqrt{\frac{4(d-2)}{\chi(q-1)}} \tan\left(2 c\tanh^{-1}
(\omega^{q-1}/r^{q-1})
\right) dt\wedge \cdots \wedge dx_p$. Unlike in the previous case, 
the solutions
now depend on two parameters $\omega$ and $\delta$. These solutions 
are new and have not been considered before. They are quite unusual
because of the presence of the periodic function and they are
not well-defined everywhere in $r$ and have possible singularities at
$2 c \tanh^{-1} (\omega^{q - 1}/r^{q - 1}) = n \pi + \pi/2$ with 
$n$ an integer. These singularities are not enclosed by the
corresponding event horizons, therefore naked. Our past experience 
\cite{jkkm,mt} tells that such singularities indicate the presence of an
external source. We hope to understand the nature of the singularities
and the associated issues better elsewhere.  These solutions 
have actually very similar structure as the solutions obtained below with the
non-extremality function $G_+(r)$ (whose singularity structure are discussed
below eq.(2.41)) and so, we will not elaborate further on these solutions here.

Having described the solutions with the non-extremality function $G_-(r)$,
we now discuss the solutions with the other non-extremality function $G_+(r)$
given in (2.18). It is clear that by letting $\omega^{q-1}$ purely imaginary
i.e. $\omega^{q-1} \to i \tilde{\omega}^{q-1}$, $G_-(r) \to G_+(r)$. In other
words the solutions with the non-extremality function $G_+(r)$ can be 
obtained from those with $G_-(r)$ by substituting $\omega^{q-1} = i 
\tilde{\omega}^{q-1}$, where $\tilde{\omega}^{q-1}$ is real. Now following
the previous solution, we find from (2.20) that since the harmonic functions
$H_2(r)$ and $\tilde{H}_2(r)$ are not real ($\ln(H_2/\tilde{H}_2) = 2i 
\tan^{-1}(\tilde{\omega}^{q-1}/r^{q-1})$, (purely imaginary)), so for the
dilaton to remain real $\delta$ must be purely imaginary. Let us put $\delta 
= -i \tilde{\delta}$, where $\tilde{\delta}$ is real. Then the dilaton 
is given
as,
\be
\phi = \frac{a(d-2)}{q-1} B - i\tilde{\delta} \ln \frac{H_2}{\tilde{H}_2}
= \frac{a(d-2)}{q-1} B + 2\tilde{\delta} \tan^{-1} \frac{\tilde{\omega}^{q-1}}
{r^{q-1}}
\ee
Let us also put $\alpha = - i\tilde{\a}$ and $\b = -i \tilde{\b}$, however,
$\tilde{\a}$ and $\tilde{\b}$ are not real in general. If we substitute these
in $F_1$ given by (2.23) it becomes,
\be
F_1 \to F_2 = \cosh^2\theta e^{2\tilde{\a}\tan^{-1}\frac{\tilde{\omega}^{q-1}}
{r^{q-1}}} - \sinh^2\theta e^{-2\tilde{\b}\tan^{-1}\frac{\tilde{\omega}^{q-1}}
{r^{q-1}}}
\ee
Now substituting $e^B=F_2^{\gamma}$, in (2.22) we find for consistency
\be
\gamma\chi = -2, \qquad \tilde{\a} - \tilde{\b} = a \tilde{\d}, \qquad
b = \sqrt{\frac{4 (q-1) (d-2)}{\chi}} (\tilde{\alpha} + \tilde{\b}) 
\tilde{\omega}^{q-1} \sinh 2\theta
\ee
Also, the consistency of equation of motion (2.12) yields a relation among
the parameters as,
\be
\frac{1}{2} \tilde{\d}^2 + \frac{2\tilde{\a}(\tilde{\a}-a\tilde{\d})(d-2)}
{\chi(q-1)} = - \frac{q}{q-1}
\ee
From (2.37) we determine $\tilde{\a}$ and $\tilde{\b}$ in terms of 
$\tilde{\d}$ as,
\bea
\tilde{\a} &=& i\tilde{c} + \frac{a\tilde{\d}}{2},\qquad\qquad
\tilde{\b}\,\, = \,\, i\tilde{c} - \frac{a\tilde{\d}}{2}\nn
{\rm with} \quad \tilde{c} &=& \sqrt{\frac{\chi q}{2(d-2)} + \frac{\tilde{\d}^2}{4}
\left(\frac{\chi(q-1)}{d-2} - a^2\right)}
\eea
Note that $\tilde c$ in the above is real. Thus we find that both $\tilde{\a}$ and
$\tilde{\b}$ are complex. But $\tilde{\a} + \tilde{\b} = 2i\tilde c$ is purely
imaginary. Thus from the last relation of (2.36) we find that for $b$ to
remain real and positive $\theta$ must be purely imaginary i.e. $\theta =
-i \tilde{\theta}$, where $\tilde{\theta}$ is real. Substituting this in $F_2$
(eq.(2.35)) we find that $F_2$ will remain real for $\tilde{\theta} = \pi/4$
only and in that csae $F_2$ becomes,
\be
F_2 = e^{a\tilde{\d}\tan^{-1}\frac{\tilde{\omega}^{q-1}}{r^{q-1}}}\,\, 
\cos\left(2\tilde c \tan^{-1} \frac{\tilde{\omega}^{q-1}}{r^{q-1}}\right)
\ee
The solutions in this case have precisely the same form as
in eq.(2.30) with $F_1$ replaced by $F_2$ (given in eq.(2.39)) and $H_1$,
$\tilde{H}_1$ replaced by $H_2$, $\tilde{H}_2$ and $\delta=-i\tilde{\d}$. 
The complete
solutions with $G_+(r)$ as the non-extremality function therefore are,
\bea
ds^2 &=& F_2^{\frac{4(p+1)}{(q-1)\chi}} \left(1+\frac{\tilde{\omega}^{2(q-1)}}
{r^{2(q-1)}}\right)^{\frac{2}{q-1}}
\left(dr^2 + r^2
d\Omega_{d-p-2}^2\right) + F_2^{-\frac{4}{\chi}}\left(-dt^2 + dx_1^2 + \cdots +
dx_{p}^2\right)\nn
e^{2\phi} &=& e^{\frac{8\tilde{\d}(p+1)}{\chi}\tan^{-1}
\frac{\tilde{\omega}^{q-1}}{r^{q-1}}}\left[\cos
\left(2\tilde c \tan^{-1}\frac{\tilde{\omega}^{q-1}}{r^{q-1}}
\right)\right]^{-\frac{4a(d-2)}{(q-1)\chi}}\nn
F_{[q]} &=& b\,\,{\rm Vol}(\Omega_{d-p-2})
\eea
For the electrical solutions, the gauge field can be obtained from (2.32) as,
\be
A_{[p+1]} = \sqrt{\frac{4(d-2)}{\chi(q-1)}} \tan\left(2\tilde c \tan^{-1}
\frac{\tilde{\omega}^{q-1}}{r^{q-1}}\right) dt \wedge \ldots \wedge dx_p
\ee
We thus find that the solutions in this case are parameterized by only two 
parameters $\tilde{\omega}$ and $\tilde{\d}$. Note that the parameter
$\tilde{\d}$ can be arbitrarily large as can be seen from eq.(2.38). This
is important to obtain the BPS limits of these solutions and will be
discussed in section 3. 
It should be noted that for both $G_-(r)$ and
$G_+(r)$ in eq.(2.8), the $p$-brane solutions we obtain are not regular 
for all $r$ between 0 and $\infty$. In fact for $G_-(r)$ with the case $(i)$
given in (2.28), the 
three parameter solutions
are not well-defined between $r=0$ and $r=\omega$ and at $r=0,\,\omega$ there
are singularities. But for $r>\omega$, the solutions are regular. By
definition, we need in general $F_1$ for $G_- (r)$ to
be positive so that the corresponding configuration is
well-defined\footnote{There might exist some possibilities that this
requirement can be relaxed. We will not discuss these in this
paper.}. This can be achieved in the above case if $r > \omega$.   On the
other hand for $G_- (r)$ with the case $(ii)$ given in eq. (2.28),
 this is not enough because of the presence of the
cosine function. We need in addition the range of validity for $r$ to be
outside of the following:
\be
\frac{1+4n}{2} \pi \,\,\leq \,\,2 c \tanh^{-1}\frac{\omega^{q-1}}
{r^{q-1}}\,\,
\leq\,\, \frac{3+4n}{2} \pi, \quad {\rm for} \quad n=0,1,2,\ldots
\ee
In other words the solutions for this case are well-defined if $r$ satisfy the 
conditions 
\bea
0 &<& 2 c \tanh^{-1}\frac{\omega^{q-1}}{r^{q-1}} \,\,\,<\,\,\,
\pi/2\nn
\frac{3+4n}{2} \pi &<& 2 c \tanh^{-1}\frac{\omega^{q-1}}{r^{q-1}}
\,\,\,<\,\,\, \frac{5+4n}{2}\pi, \quad {\rm for} \quad n=0,1,2,\ldots
\eea
Note that since $c > 0$ and 
$0 < \tanh^{-1}\frac{\omega^{q-1}}{r^{q-1}} < \infty$ as $r >
\omega$, we therefore have $r > \omega /[\tanh
(\pi/4c)]^{1/(q - 1)}$ from the first equation of (2.43)  and 
\be
\frac{ \omega}{[\tanh(\frac{5 + 4n}{4 c} \pi)]^{1/(q - 1)}} < r <
  \frac{ \omega}{[\tanh(\frac{3 + 4n}{4 c} \pi)]^{1/(q - 1)}}
\ee
from the second equation.
The above implies that apart from the first case, the solutions are
well-defined only in a finite region of $r$ determined by $ c$
and integer $n$.  
For the case of $G_+ (r)$ and to have a positive $F_2$, we find, 
from eqs.(2.39) and (2.41), that $r$ must lie  outside the region 
given by (2.42)
 but with $ c \tanh^{-1}\frac{\omega^{q-1}}
{r^{q-1}}$ replaced by $\tilde c \tan^{-1}\frac{\tilde{\omega}^{q-1}}
{r^{q-1}}$. We have the similar replacement in eq. (2.43) for the
present case. By the same reasoning as discussed in footnote 3, one can
show now $2 \tilde c > 1$. Note also by definition $ 0 < \tan^{-1} (\tilde
\omega^{q - 1}/r^{q - 1}) < \pi/2$, since $r > 0$. The analogous
equation of the first one in (2.43) in the present case gives $r > 
\omega/[\tan\,(\pi/4 \tilde c)]^{1/(q - 1)}$ and the remaining case is
subtle and needs to be considered carefully as follows: (a) if $(3 +
4n)/2\tilde c > 1$, then no solutions are allowed since the lower bound already
exceeds $\pi/2$; (b) if $(3 +
4n)/2\tilde c < 1$ but $(5 + 4n)/2\tilde c > 1$ (i.e, $\tilde c/2 - 3/4
> n > \tilde c /2 - 5/4$), the allowed region of validity for $r$ is
 $\omega/[\tan\,(3 + 4n)\pi/(4 \tilde c)]^{1/(q - 1)} >    r > 
\omega/[\tan\,\pi/(4 \tilde c)]^{1/(q - 1)}$; (c) if $(5 + 4n)/(2 \tilde
c) < 1$, the allowed region of validity is 
$\omega/[\tan\,(3 + 4n)\pi/(4 \tilde c)]^{1/(q - 1)} > r >
\omega/[\tan\,(5 + 4n)\pi/(4 \tilde c)]^{1/(q - 1)}$.  
 For each of the cases discussed above, a possible singularity can occur on
the border of the region of validity for $r$. These singularities are
naked in nature and therefore indicate the presence of an external
source as mentioned earlier.

\vspace{.5cm}

\noindent{\bf $p=-1$-brane or D-instanton}

\vspace{.2cm}

So far we have discussed various static, non-supersymmetric $p$-brane
solutions in type II supergravities which in principle include $p=-1$
case. However, since the solutions are different in some respects we
discuss this case separately. First, we note from the condition (2.5)
that in this case there is no $B(r)$ in the metric. Also, since $q=d-p-2
=d-1$, so, from (2.8) we have
\be
A(r) = \frac{1}{d-2}\ln G(r)
\ee
where as before $G(r)$ can take two forms 
\bea
G_-(r) &=& 1 - \frac{\omega^{2(d-2)}}{r^{2(d-2)}} = \left(1+\frac{\omega^{d-2}}
{r^{d-2}}\right)\left(1-\frac{\omega^{d-2}}{r^{d-2}}\right) =
H_1 (r) {\tilde H}_1 (r)\nn
G_+(r) &=& 1 + \frac{\tilde{\omega}^{2(d-2)}}{r^{2(d-2)}} = \left(1+\frac{i
\tilde{\omega}^{d-2}}
{r^{d-2}}\right)\left(1-\frac{i\tilde{\omega}^{d-2}}{r^{d-2}}\right) =
H_2 (r) {\tilde H}_2 (r)
\eea
The above form of $A(r)$ is consistent with the equations of motion (2.14).
The equation of motion (2.13) is absent and we rewrite the other two
equations of motion (2.15) and (2.12) with $p=-1$, $q=d-2$ as,
\bea
\phi'' + \frac{q}{r} \phi' + \frac{G'}{G}\phi' - \frac{ab^2}{2}
\frac{e^{a\phi}}{G^2 r^{2(d-1)}} &=& 0\\
(d-1)\left(A''+\frac{A'}{r}\right) + \frac{1}{2} \phi'^2 - \frac{b^2}{2} 
\frac{e^{a\phi}}{G^2 r^{2(d-1)}} &=& 0
\eea
As before the equation (2.47) can be solved for the non-extremality function
$G_-(r)$ with an ansatz for $e^{\phi}$
as,
\be
e^{\phi} = F_1^\nu
\ee
with
\be
F_1 = \left[\cosh^2\theta \left(\frac{H_1}{\tilde {H}_1}\right)^\alpha -
\sinh^2\theta
\left(\frac{\tilde {H}_1}{H_1}\right)^\beta\right], \quad
H_1 = 1 + \frac{\omega^{d-2}}{r^{d-2}}, \quad \tilde{H}_1 = 
1 - \frac{\omega^{d-2}}
{r^{d-2}}
\ee
where $\alpha$, $\beta$ and $\theta$ are constant parameters. The parameter
$\nu$ can be determined by substituting the above in (2.47) and we get,
\be
\nu=-\frac{2}{a}, \quad \alpha=\beta, \quad {\rm and} \quad b=\frac{4\alpha}{a}
(d-2) \omega^{d-2} \sinh 2\theta
\ee
Also, (2.48) determines the value of $\alpha$ as,
\be
\alpha = a \sqrt{\frac{(d-1)}{2(d-2)}}
\ee
So, the full non-supersymmetric D-instanton solution has the form,
\bea
ds^2 &=& \left(H_1 \tilde{H}_1\right)^{\frac{2}{d-2}}\left(dr^2 + r^2 
d\Omega_{d-1}^2\right)\nn 
e^{2\phi} &=& \left[\cosh^2\theta \left(\frac{H_1}{\tilde {H}_1}
\right)^\alpha - \sinh^2\theta
\left(\frac{H_1}{\tilde{H}_1}\right)^{-\alpha}\right]^{-\frac{4}{a}}\nn
F_{[d-1]} &=& b {\rm Vol}(\Omega_{d-1})
\eea
The scalar field for the corresponding electrically charged solution has the
form
\be
A_{[0]} = \frac{2i}{a} \sinh\theta \cosh\theta\left(\frac{C_1}{F_1}\right)
\ee
where $C_1 = \left(\frac{H_1}{\tilde{H}_1}\right)^\alpha
- \left(\frac{H_1}{\tilde{H}_1}\right)^{-\alpha}$. 
Note here that the scalar $A_{[0]}$ is purely imaginary because we 
have used the
same definition of Hodge duality as was used for the other $p$-brane solutions
in (2.31). However, since for the instanton solution we need to go to 
Euclidean coordinate or equivalently change the scalar $A_{[0]}
=i{\tilde A}_{[0]}$ 
\cite{ggp} (this
changes the sign on the kinetic energy term of the scalar) and so for the
instanton solution the scalar will  take the form,
\be
\tilde{A}_{[0]} = \frac{2}{a} \sinh\theta 
\cosh\theta\left(\frac{C_1}{F_1}\right)
\ee
The eqs.(2.53), (2.55)
represent the D-instanton solution for the non-extremality function $G_-(r)$.
Note that here the solution is charaterized by two parameters $\omega$ and
$\theta$ in contrast to the three parameters for other values of $p$.
Also since here $\alpha$ in (2.51) is real there is no solution analogous
to the two parameter solution with $G_-(r)$ for the case of D-instanton. Now as
before we will obtain the solution with the non-extremality function $G_+(r)$
by substituting in (2.53) and (2.55) the following
\be
\omega^{d-2} \to i \tilde{\omega}^{d-2}, \quad H_1,\,\,\tilde{H}_1 \to
H_2,\,\,\tilde{H}_2, \quad \theta \to -i \tilde{\theta} = -i\pi/4
\ee
then the solutions take the forms,
\bea
ds^2 &=& \left(1+\frac{\tilde{\omega}^{2(d-2)}}{r^{2(d-2)}}
\right)^{\frac{2}{d-2}}
\left(dr^2 + r^2 
d\Omega_{d-1}^2\right)\nn 
e^{2\phi} &=& \left[\cos\left(2\alpha \tan^{-1}\frac{\tilde{\omega}^{d-2}}
{r^{d-2}}\right)\right]^{-\frac{4}{a}}\nn
F_{[d-1]} &=& b {\rm Vol}(\Omega_{d-1})
\eea
and for the electrically charged solution
\be
A_{[0]} = \frac{2}{a} \tan\left(2\alpha\tan^{-1}\frac{\tilde{\omega}^{d-2}}
{r^{d-2}}\right)
\ee
The singularity structures of the D-instanton solutions remain exactly the
same as we have discussed earlier for other $p$-branes with $\tilde c$ replaced
by $\alpha$ and so we will not
repeat it here. We just mention that in $d=10$ and for the usual dilaton
coupling $a=(p-3)/2=-2$, the instanton solution (2.57) and (2.58) is not 
well-defined as $r \to 0$ (given our previous discussion for $p \ne -1$,
we know that $r$ cannot be allowed to approach zero). 
A general non-supersymmetric D-instanton solution 
carrying 
electric charges of an SL(2,R) symmetry has been given recently in 
\cite{bcgrv}.
The singularity structure of these solutions and how to resolve them in some
cases have been discussed there. In the next section, we will see, among
other things, how BPS $p$-branes can be recovered from these non-supersymmetric
$p$-branes by scaling the parameters in appropriate ways.

\sect{Discussion on some aspects of the solutions}

In this section we will mainly discuss two aspects of the solutions 
obtained in the previous section, namely, how a subclass of the solutions 
can be 
regarded as the interpolating
solutions between the chargeless  $p$-brane--anti $p$-brane system  
and the usual 
BPS $p$-branes\footnote{In the context of open string tachyon condensation
the non-BPS $(p+1)$-brane on the tachyonic kink goes over to a
configuration which can be identified as BPS $p$-brane \cite{asone}. So,
the case we are discussing here is not quite the same. However, the above
process can be understood from a delocalized, non-supersymmetric $p$-brane
solution and will be discussed elsewhere \cite{lr}.} and
then we point out how the Wick rotation of these solutions lead to the
time-dependent solutions of ref.\cite{br}. But before that we clarify the
relations between the solutions obtained in section 2 and those given
in ref.\cite{zz}.

The solutions obtained by Zhou and Zhu in \cite{zz} are the generalized black
$p$-branes in $d$ dimensions and is given in eqs.(112) -- (124) of their
paper. The non-supersymmetric $p$-brane solutions in this paper correspond
to $c_2=0$ of ref.\cite{zz}. Also we note that we should make the following
identifications to compare the two solutions,
\bea
D &\equiv & d, \qquad d \equiv p+1, \qquad {\tilde d} \equiv
q-1,\nn
\Delta &\equiv & \frac{(q-1)\chi}{d-2}, \qquad r_0 \equiv \omega
\eea
where in the above we have kept the symbols used by Zhou and Zhu on the
left hand side and the symbols used in section 2 on the right.
With these identifications, we find
\be
h(r) = \ln\frac{\tilde {H}_1(r)}{H_1(r)}, \qquad \xi(r) =
\ln H_1(r) \tilde {H}_1(r) = \ln G_-(r)
\ee
We therefore have,
\be
\cosh({\tilde k} h(r)) + c_3 \sinh({\tilde k}h(r)) =
\frac{1}{2}(c_3+1) \left(\frac{H_1}{\tilde {H}_1}\right)^{-\tilde k} -
\frac{1}{2} (c_3-1) \left(\frac{\tilde {H}_1}{H_1}\right)^{-\tilde k}
\ee
Using these relations we simplify the metric and the prefactors multiplying
the longitudinal as well as the transverse parts of the brane from eqs.(120)
and (121) of ref.\cite{zz} and identify with $F_1^{-4/\chi}$ and 
$(H_1{\tilde {H}_1})^
{2/(q-1)} F_1^{4(p+1)/(q-1)\chi}$ respectively. We thus obtain,
\be
F_1 \equiv
\left(\frac{c_3+1}{2}\right) \left(\frac{H_1}{\tilde {H}_1}
\right)^{\frac{ac_1}{2}
-{\tilde k}} -
\left(\frac{c_3-1}{2}\right) \left(\frac{\tilde {H}_1}{H_1}\right)^{-\frac{
ac_1}{2}
-{\tilde k}}
\ee
Comparing with the form of $F_1$ in eq.(2.23),
the parameters in the two solutions can be related as,
\bea
c_3 &=& \cosh2\theta\nn
c_1 &=& \delta\nn
{\tilde k} &=& -\frac{1}{2}(\alpha+\beta)
\eea
With (3.5), the parameter relation given in (117), (118) of ref.\cite{zz} 
reduce
to eq.(2.26) i.e.,
\be
-\frac{4{\tilde k}^2}{\Delta} = c_1^2 - \frac{a^2 c_1^2}{\Delta} -
\frac{2({\tilde d} +1)}{\tilde d} \quad \Rightarrow \quad \frac{1}{2}
\delta^2 + \frac{2\alpha(\alpha-a\delta)(d-2)}{\chi(q-1)} = \frac{q}{q-1}
\ee
We have thus clarified the relation of the solutions obtained in ref.\cite{zz}
with the non-supersymmetric $p$-brane solutions we have obtained in eqs.(2.30),
 (2.32). We would like to point out that since ref.\cite{zz} contains only the
three parameter solutions, we have clarified their relations with the
three parameter solutions we have obtaned in section 2 with $G_-(r)$ as the
non-extremality function. Also the D-instanton solution was not given in 
\cite{zz}.

Let us now discuss how we can regard the solutions (2.30), (2.32) as
interpolating solutions between the chargeless  D$p$--anti D$p$ system
  and the usual BPS
D$p$-branes. We will also discuss a similar interpretation as interpolating
solution for the case of D-instanton solution (2.53), (2.55)\footnote{Note
that this interpretation holds only for the three parameter solutions with the
non-extremality function $G_-(r)$ and not for the two parameter solution
with $G_-(r)$ and also for $G_+(r)$ (even though there exist
BPS limits in these cases) and will be mentioned
later.}. Note that like the chargeless D$p$-antiD$p$ system 
non-BPS D$p$-branes also have net RR charge zero. The reason is the non-BPS 
D$p$-branes of even (odd) dimensionalities exist in type IIB (IIA) 
superstring theory as opposed to their
BPS counterpart of odd (even) dimensionalities in the same theory\cite{asone}. 
However, we know that type IIA (IIB) string 
theory contains
odd (even) form RR gauge fields and the D-branes are charged under these
gauge fields. Since a D$p$-brane couples to a $(p+1)$-form gauge field, the
charged D$p$-branes in type IIA (IIB) theory must be of even (odd)
dimensionality. However, since the non-BPS D$p$-branes are of opposite
i.e. odd (even) dimensionalities in IIA (IIB) theories, they must be
chargeless. So, what we want to emphasize here is that from an isolated
supergravity solution it is not possible to distinguish a non-BPS D$p$-brane
from a D$p$-antiD$p$ system of zero RR charge \cite{bmo,divec}. But since the
general solutions (2.30), (2.32) interpolate (as we will show) between two 
solutions belonging to the same theory, so, if one solution is the BPS
D$p$-brane, the other one can not be non-BPS D$p$-brane (since they do not 
belong to the same theory) and has to be coincident D$p$-antiD$p$-brane
system with zero net charge.
  
We note from the solutions (2.30) that $F_{[q]}=0$ implies $b=0$. Also from
the third relation of (2.25) we find $b=0$ implies $\theta=0$. Therefore
the function $F_1$ in (2.23) reduces to
\be
F_1 = \left(\frac{H_1}{\tilde {H}_1}\right)^\alpha
\ee
The complete brane--anti brane solutions can then be seen from
(2.30), (2.32) to take the forms,
\bea
ds^2 &=& (H_1 {\tilde {H}_1})^{\frac{2}{q-1}}\left(\frac{H_1}{\tilde {H}_1}
\right)^
{\frac{4(p+1)}{(q-1)\chi} \alpha}\left(dr^2 + r^2 d\Omega_{d-p-2}^2\right)
+ \left(\frac{H_1}{\tilde {H}_1}\right)^{-\frac{4}{\chi}\alpha}
\left(-dt^2+dx_1^2
+\cdots + dx_p^2\right)\nn
e^{2\phi} &=& \left(\frac{H_1}{\tilde {H}_1}\right)^{-\frac{4a(d-2)}{(q-1)\chi}
\alpha + 2\delta}\nn
F_{[q]} &=& 0, \qquad A_{[p+1]}\,\,\,=\,\,\,0
\eea
where the parameters $\alpha$ and $\delta$ are related by the first relation
of eq.(2.27). These solutions are now characterized by two parameters $\omega$
and $\delta$ and in analogy with the arguments given in 
refs.\cite{bmo,luroy}, the
parameters would presumably be related to the mass and the tachyon vev of the
underlying unstable brane--anti brane system. In $d=10$, $\chi = 32/(7-p)$, $a=(p-3)/2$, the solutions
(3.8) simplify to,
\bea
ds^2 &=& (H_1 {\tilde {H}_1})^{\frac{2}{7-p}}\left(\frac{H_1}{\tilde {H}_1}
\right)^
{\frac{(p+1)}{8} \alpha}\left(dr^2 + r^2 d\Omega_{8-p}^2\right)
+ \left(\frac{H_1}{\tilde {H}_1}\right)^{-\frac{7-p}{8}\alpha}
\left(-dt^2+dx_1^2
+\cdots + dx_p^2\right)\nn
e^{2\phi} &=& \left(\frac{H_1}{\tilde {H}_1}\right)^{\frac{3-p}{2}
\alpha + 2\delta}\nn
F_{[8-p]} &=& 0, \qquad A_{[p+1]}\,\,\,=\,\,\,0
\eea
with,
\be
\alpha = \sqrt{\frac{2(8-p)
}{(7-p)} - \frac{(7-p)(p+1)}{16}\delta^2}
+ \frac{(p-3)\delta}{2}
\ee
This is exactly the same  supergravity solutions obtained in
refs.\cite{bmo,divec} even though our interpretation here is different. 
For the case of D-instanton solution the corresponding
brane--anti brane solution can be obtained by setting as before $b=\theta=0$ in
(2.53), (2.55) and the solutions take the form,
\bea
ds^2 &=& \left(H_1 \tilde{H}_1\right)^{\frac{2}{d-2}}\left(dr^2 + r^2 
d\Omega_{d-1}^2\right)\nn 
e^{2\phi} &=& \left(\frac{H_1}{\tilde {H}_1}
\right)^{-\frac{4\alpha}{a}}\nn
F_{[d-1]} &=& 0, \qquad A_{[0]} = 0 
\eea
where $\alpha = a \sqrt{(d-1)/[2(d-2)]}$. In $d=10$, they have the forms
\be
ds^2 = \left(H_1 \tilde{H}_1\right)^{\frac{1}{4}}\left(dr^2 + r^2 
d\Omega_{9}^2\right), \quad 
e^{2\phi} = \left(\frac{H_1}{\tilde {H}_1}
\right)^{-3}, \quad
F_{[9]} = 0, \quad A_{[0]} = 0 
\ee
where $H_1 = 1 + \omega^8/r^8$, $\tilde{H}_1 = 1 - \omega^8/r^8$. We would
like to point out that for the solutions with $G_+(r)$ as the 
non-extremality function, there are no non-trivial chargeless solutions. 
This is because in this case $\tilde{\theta}=\pi/4$ as we noted before and
so, for the charge to vanish $\tilde{\omega}$ has to vanish for all $p$
from $-1$ to 6 (see eq.(2.36)). So, the chargeless sloution in this case
is trivial i.e. the flat space. (This conclusion also holds for the two
parameter solutions with $G_-(r)$ as the non-extremality function.)

Now we will see how BPS $p$-branes can be obtained from the same solutions
(2.30), (2.32). We first note from the third relation in (2.25) that if
$(\alpha+\beta) \omega^{q-1} \to \epsilon {\bar \omega}^{q-1}$ and $\sinh
2\theta \to \epsilon^{-1}$, where $(\alpha+\beta)$ = finite, $\epsilon$ is
a dimensionless parameter with $\epsilon \to 0$ and ${\bar \omega}^{q-1}
= \sqrt{\frac{b^2\chi}{4(q-1)(d-2)}}$, then this relation reduces to the
usual mass-charge relation of the magnetically charged BPS $p$-branes in
$d$ dimensions. (Note that $b$ = fixed in this case.) Now it is clear that
since $\omega^{q-1} \to \epsilon\, {\bar \omega}^{q-1}/(\alpha+\beta)$,
both $H_1(r)$ and ${\tilde {H}_1}(r) \to 1$. On the other hand we 
find from (2.23)
\bea
F_1 &=& \cosh^2\theta \left(\frac{H_1}{\tilde {H}_1}\right)^\alpha - 
\sinh^2\theta
\left(\frac{\tilde {H}_1}{H_1}\right)^\beta\nn
&=& 1 + \frac{\left[(\alpha+\beta)\cosh2\theta + (\alpha-\beta)\right]
\omega^{q-1}}{r^{q-1}}\nn
&\to & 1+\frac{{\bar \omega}^{q-1}}{r^{q-1}}\,\,\,=\,\,\, {\bar {H}_1}(r)
\eea
Note here that the parameters $\alpha$, $\beta$ (or $\delta$) remain arbitrary
but they do not appear in the solutions. In these limits the gauge field in
(2.32) reduces to
\be
A_{[p+1]} = \sqrt{\frac{4(d-2)}{\chi(q-1)}} \left(1-{\bar {H}_1}^{-1}\right)
dt \wedge dx_1 \wedge \cdots \wedge dx_p
\ee
We thus recover the BPS $p$-brane solutions in $d$ dimensions from the
solutions (2.30), (2.32) in the limit
\bea
(\alpha+\beta) \omega^{q-1} &\to & \epsilon\,{\bar \omega}^{q-1}\nn
\sinh2\theta & \to & \epsilon^{-1}
\eea
with $\epsilon \to 0$ and $(\alpha+\beta)$ = finite. In $d=10$, $\chi
= 32/(7-p)$ and using the above relations, the solutions (2.30), (2.32)
take the forms,
\bea
ds^2 &=& {\bar {H}_1}^{\frac{p+1}{8}}
\left(dr^2 + r^2 d\Omega_{8-p}^2\right)
+ {\bar {H}_1}^{-\frac{7-p}{8}}\left(-dt^2+dx_1^2
+\cdots + dx_p^2\right)\nn
e^{2\phi} &=& {\bar {H}_1}^{\frac{3-p}{2}
}\nn
F_{[8-p]} &=& b {\rm Vol}(\Omega_{8-p})
\eea
for the magnetic brane and for the electric brane,
\be
A_{[p+1]} = \left(1-{\bar {H}_1}^{-1}(r)\right) dt \wedge dx_1 \wedge 
\cdots \wedge
dx_p
\ee
These are precisely the BPS magnetic and electric $p$-brane solutions in
$d=10$.

In deriving the BPS $p$-branes from the non-supersymmetric $p$-branes, we
have taken the limits (3.15), where we also kept $\alpha+\beta$ = finite.
However, we can also recover the BPS $p$-branes by taking another limit
as,
\bea
\alpha + \beta &\to & \epsilon^{\frac{1}{2}}\nn
\omega^{q-1} &\to & \epsilon^{\frac{1}{2}}\,{\bar \omega}^{q-1}\nn
\sinh2\theta &\to & \epsilon^{-1}
\eea
where ${\bar \omega}$ is as defined before. We note that even in this case
$H_1(r) \to 1$ and ${\tilde {H}_1}(r) \to 1$ and $F_1 \to {\bar {H}_1}(r) 
= 1 + \frac{
{\bar \omega}^{q-1}}{r^{q-1}}$ as before and
$A_{[p+1]} \to \sqrt{\frac{4(d-2)}{\chi(q-1)}}\left(1-{\bar 
{H}_1}^{-1}(r)\right)
dt \wedge dx_1 \wedge \cdots \wedge dx_p =
\left(1-{\bar {H}_1}^{-1}(r)\right)
dt \wedge dx_1 \wedge \cdots \wedge dx_p$ in $d=10$. But unlike in the previous
case where $\alpha$, $\beta$ (or $\delta$) remains arbitrary, here they
scale. Since $\alpha+\beta = -2{\tilde k} = 2 \sqrt{\frac{\chi q}{2(d-2)}
- \frac{\delta^2}{4}\left(\frac{\chi(q-1)}{d-2}-a^2\right)}$, so, $\alpha
+ \beta \to \epsilon^{1/2}$ implies
\be
|\delta| \to \left[\frac{\chi q}{(q-1)(p+1)}\right]^{\frac{1}{2}}
- \frac{\epsilon\, (d-2)}{4[\chi q (q-1)(p+1)]^{\frac{1}{2}}}
\ee
In $d=10$, this limit has been taken in ref.\cite{bmo} to recover 
BPS D$p$-branes
from the solutions given in \cite{zz}\footnote{Note here that the BPS 
limits we have 
discussed hold only for the solution when $|\d|$ is bounded by case $(i)$
of (2.28). However, for case $(ii)$ there is also a BPS limit analogous
to eq.(3.23) given below and the BPS solution in this case take the forms
very similar to eq.(3.25). So, we do not elaborate the BPS limits in this case
whose meanings are also not clear to us.}. 

For the case of D-instanton solution (2.53), (2.55) a limit similar to (3.15)
can be taken. However, since for this case $\alpha$ is fixed, there is no 
limit similar to (3.18). So, for $0\leq p \leq 6$, there are two distinct
ways in which BPS $p$-brane solutions can be recovered by scaling the 
parameters of the non-supersymmetric $p$-brane solutions, but for $p=-1$ or for
D-instanton there is only one way the BPS solution can be recovered. Let us
indicate how this is done for $p=-1$ case. We note from the expression of 
$F_1$ in (2.50) that with the following scaling of the parameters,
\bea
2\alpha \omega^{d-2} &\to& \epsilon \bar{\omega}^{d-2}\nn
\sinh\theta &\to& \epsilon^{-1}
\eea
where $\epsilon \to 0$, $F_1$ reduces to,
\be
F_1 \to \bar{H}_1 = 1 + \frac{\bar{\omega}^{d-2}}{r^{d-2}}
\ee
Then the solutions (2.53), (2.55) reduce to
\bea
ds^2 &=& \left(dr^2 + r^2 
d\Omega_{d-1}^2\right)\nn 
e^{2\phi} &=& (\bar{H}_1)^{-\frac{4}{a}}\nn
F_{[d-1]} &=& b {\rm Vol}(\Omega_{d-1}), \qquad A_{[0]} = \frac{2}{a}\left(1-
\bar{H}_1^{-1}\right) 
\eea
In $d=10$ and $a=-2$, this is precisely the D-instanton solution obtained
in ref.\cite{ggp}. This is a regular solution where the metric has the wormhole
geometry in the string frame.

We have therefore shown how the
non-supersymmetric $p$-brane solutions (2.30), (2.32) and (2.53), (2.55) can 
be regarded as
interpolating solutions from brane--anti brane solutions (for $\theta \to 0$) to BPS
$p$-branes (for $\theta \to \infty$ and keeping $b$ fixed). For the case
of $0\leq p \leq 6$, the BPS solutions were obtained in two different ways
whereas for $p=-1$ it was obtained in only one way. 
We would also like to
point out that in recovering BPS $p$-brane solutions $\sinh 2\theta$ has to
go to infinity and this is not possible for trigonometric function which
appears in the case of corresponding time-dependent solutions. This is
consistent with the fact that for time-dependent case there are no real
BPS solutions in type II supergravities.

For the solutions (2.40)
with $G_+(r)$ as the non-extremality function we mentioned before that there is
no non-trivial chargeless solutions analogous to brane--anti brane
systems in this 
case. However, it is possible to obatin BPS solutions by scaling the 
parameters in appropriate ways. Let us indicate how this can be done from 
(2.40). We scale the parameters as follows,
\bea
\tilde{\omega}^{q-1} &\to& \tilde{\e} \bar{\omega}^{q-1}\nn
\tilde{\d} &\to& \tilde{\e}^{-1}
\eea
where $\tilde{\e}$ is a dimensionless parameter with $\tilde{\e} \to 0$
and $\bar{\omega}^{q-1}$ = fixed.
Note that with this scaling $G_+(r) \to 1$ and the condition (2.8) reduces
to the supersymmetry condition. The function $F_2$ in (2.39) takes the form,
\be
F_2 \to \bar{F}_2 = e^{a\frac{\bar{\omega}^{q-1}}{r^{q-1}}} \cos\left(
2\bar{c} \frac{\bar{\omega}^{q-1}}{r^{q-1}}\right)
\ee
where $\bar{c} = \sqrt{\frac{(p+1)(q-1)}{2(d-2)}}$. It is clear that 
since $F_2$ 
contains a periodic function of $r$, it can not be reduced to the usual harmonic 
function of a BPS $p$-brane. The complete BPS solutions in this case have the
forms, 
\bea
ds^2 &=& \bar{F}_2^{\frac{4(p+1)}{(q-1)\chi}} 
\left(dr^2 + r^2
d\Omega_{d-p-2}^2\right) + \bar{F}_2^{-\frac{4}{\chi}}
\left(-dt^2 + dx_1^2 + \cdots +
dx_{p}^2\right)\nn
e^{2\phi} &=& e^{4\frac{\bar{\omega}^{q-1}}{r^{q-1}}}
\bar{F}_2^{-\frac{4a(d-2)}{(q-1)\chi}}\nn
F_{[q]} &=& b\,\,{\rm Vol}(\Omega_{d-p-2})
\eea
for the magnetically charged solutions and for the electrical solutions we have
\be
A_{[p+1]} = \sqrt{\frac{4(d-2)}{\chi(q-1)}}\tan \left(2\bar{c} \frac{\bar
{\omega}^{q-1}}{r^{q-1}}\right)
\ee
These BPS solutions are not of the usual BPS $p$-brane type as they involve
periodic functions. Like the solutions (2.40) these are also not
well-defined for all $r$ between 0 and $\infty$. In fact these solutions are
well-defined inside the range of $r$ given by $r > \bar\omega/[\pi/4\bar
c]^{1/(q -1)}$ or by
\be
\frac{\bar{\omega}}{\left(\frac{5 +4n}{2\bar{c}}\pi\right)^{\frac{1}{q-1}}}
< r <
\frac{\bar{\omega}}{\left(\frac{3+4n}{2\bar{c}}\pi\right)^{\frac{1}{q-1}}},
\qquad {\rm for} \quad n=0,1,2,\ldots
\ee
and $r=0$ is excluded and there are singularities at
\be
r= \bar\omega/[\pi/4\bar c]^{1/(q -1)},\, 
\frac{\bar{\omega}}{\left(\frac{3+4n}{4 \bar{c}}\pi\right)^{\frac{1}{q-1}}}
,\,
\frac{\bar{\omega}}{\left(\frac{5 +4n}{4\bar{c}}\pi\right)^{\frac{1}{q-1}}}
\ee
So, we mention that although the solutions (2.40) have BPS limits, these are
quite unusual and therefore, (2.40) can not be interpreted as 
interpolating solutions
of p-brane--anti p-brane systems and the usual BPS $p$-branes 
as for the three parameter
solutions with
$G_-(r)$.

Next we show how by a Wick rotation on the static non-supersymmetric $p$-brane
solutions given in (2.30), (2.32) we get the time-dependent solutions
or space-like $p$-branes \cite{stro,cgg,kmp,roy,ohta} (or S$p$-branes) obtained 
in \cite{br}. Usually the 
Wick rotations on the BPS $p$-branes do
not lead to real solutions, however, in this case we get real solutions.
Let us consider the solutions (2.30), (2.32) and apply
the following Wick rotation,
\bea
r &\to & it \nn
t &\to & i x_{p+1}
\eea
along with $\omega \to i \omega$. We also write,
\be
d\Omega_{d-p-2}^2 = d\psi^2 + \sin^2\psi d\Omega_{d-p-3}^2
\ee
and make the Wick rotation $\psi \to -i\psi$. Under this change (3.29) reduces
to,
\be
d\Omega_{d-p-2}^2 = d\psi^2 + \sin^2 \psi d\Omega_{d-p-3}^2 \to
-d\psi^2 - \sinh^2\psi d\Omega_{d-p-3}^2 = -dH_{d-p-2}^2
\ee
where $dH_{d-p-2}^2$ is the line element for the $(d-p-2)$-dimensional
hyperbolic space. By further changing $\theta \to i\theta$, we find that
since
\bea
H_1(r) &=& 1 + \frac{\omega^{q-1}}{r^{q-1}}\,\,\, \to \,\,\, 1+
\frac{\omega^{q-1}}{t^{q-1}} \,\,\, =\,\,\, H_1(t)\nn
{\tilde {H}_1}(r) &=& 1 - \frac{\omega^{q-1}}{r^{q-1}}\,\,\, \to \,\,\, 1-
\frac{\omega^{q-1}}{t^{q-1}} \,\,\, =\,\,\, {\tilde {H}_1}(t)\nn
F_1(r) &=& \cosh^2\theta \left(\frac{H_1(r)}{{\tilde {H}_1}(r)}\right)^\alpha
- \sinh^2\theta
\left(\frac{{\tilde {H}_1}(r)}{H_1(r)}\right)^\beta\nn
& \to &
\cos^2\theta \left(\frac{H_1(t)}{{\tilde {H}_1}(t)}\right)^\alpha + 
\sin^2\theta
\left(\frac{{\tilde {H}_1}(t)}{H_1(t)}\right)^\beta\,\,\,=\,\,\, F_1(t)
\eea
the metric and the dilaton in (2.30) changes to,
\bea
ds^2 &=& F_1(r)^{\frac{4(p+1)}{(q-1)\chi}} (H_1(r){\tilde {H}_1}(r))^
{\frac{2}{q-1}}
\left(dr^2 + r^2
d\Omega_{d-p-2}^2\right) + F_1(r)^{-\frac{4}{\chi}}
\left(-dt^2 + \cdots +
dx_{p}^2\right)\nn
&\to & F_1(t)^{\frac{4(p+1)}{(q-1)\chi}} (H_1(t){\tilde {H}_1}(t))^
{\frac{2}{q-1}}
\left(-dt^2 + t^2
dH_{d-p-2}^2\right) + F_1(t)^{-\frac{4}{\chi}}
\left(dx_1^2 + \cdots +
dx_{p+1}^2\right)\nn
e^{2\phi} &=& F_1(r)^{-\frac{4a(d-2)}{(q-1)\chi}}
\left(\frac{H_1(r)}{{\tilde {H}_1}(r)}\right)^{2\delta}\,\,\, \to
F_1(t)^{-\frac{4a(d-2)}{(q-1)\chi}}
\left(\frac{H_1(t)}{{\tilde {H}_1}(t)}\right)^{2\delta}
\eea
Now in order to see how $F_{[q]}$ changes we first note that
\bea
{\rm Vol}(\Omega_{d-p-2}) &=& (\sin\psi)^{d-p-3} d\psi \wedge \cdots\nn
& \to & (-i)^{d-p-2} (\sinh \psi)^{d-p-3} d\psi \wedge \cdots\nn
&=& (-i)^{d-p-2} {\rm Vol}(H_{d-p-2})
\eea
where ${\rm Vol}(H_{d-p-2})$ is the volume-form of the hyperbolic space.
It is clear from (3.33) that in order to get a real solution $b$ must change
as $b \to (i)^{d-p-2} b$ and $F_{[q]}$ then changes to
\be
F_{[q]} = b {\rm Vol}(\Omega_{d-p-2}) \to b {\rm Vol}(H_{d-p-2})
\ee
It can be easily checked that the parameter relation changes as,
\be
b = \sqrt{\frac{4(q-1)(d-2)}{\chi}}(\alpha+\beta) \omega^{q-1} \sinh 2\theta
\quad\to \quad b = \sqrt{\frac{4(q-1)(d-2)}{\chi}}(\alpha+\beta) \omega^{q-1}
\sin 2\theta
\ee
With these changes the gauge field (2.32) changes to
\be
A_{[p+1]} \to \sqrt{\frac{4(d-2)}{\chi(q-1)}} \sin\theta \cos\theta
\left(\frac{C_1}{F_1}\right) dx_1\wedge \cdots \wedge dx_{p+1}
\ee
upto an overall sign. Eqs.(3.32), (3.34) -- (3.36) are precisely the form of
the time dependent solutions we obtained in \cite{br}.

Let us now look at the D-instanton solution (2.53) and (2.54)\footnote{
The reason we are using (2.54) as the solution for the scalar instead
of (2.55) is that for the case of S($-1$)-brane we do not need to go to
the Euclidean coordinate as for the D($-1$)-brane solution.} with $G_-(r)$
as the non-extremality function. By using the same trick as applied for 
other $p$-branes, we here obtain the corresponding time-dependent solution
or S($-1$)-brane solution as,
\bea
ds^2 &=& \left(1-\frac{\omega^{2(d-2)}}{t^{2(d-2)}}\right)^{\frac{2}{d-2}}
\left(-dt^2 + t^2 
dH_{d-1}^2\right)\nn 
e^{2\phi} &=& \left[\cos^2\theta \left(\frac{H_1(t)}{\tilde {H}_1(t)}
\right)^\alpha + \sin^2\theta
\left(\frac{H_1(t)}{\tilde{H}_1(t)}\right)^{-\alpha}\right]^{-\frac{4}{a}}\nn
F_{[d-1]} &=& b {\rm Vol}(H_{d-1}), \qquad A_{[0]}\,\,=\,\, 
\frac{2}{a} \sin\theta
\cos\theta \left(\frac{C_1(t)}{F_1(t)}\right)
\eea
In $d=10$ this is precisely the S$(-1)$-brane solution obtained in 
ref.\cite{kmp}.
It can be easily checked that for the $p$-brane solutions, eqs.(2.40),
(2.41) including the D-instanton solution, eqs.(2.57), (2.58) with $G_+(r)$
as the non-extremality function, there are no real time-dependent solutions
which can be obtained by Wick rotation. The reason is, in this case as we
make the Wick rotation $r \to it$ and $\tilde{\omega} \to i \tilde{\omega}$,
we note from (2.36) that $b$ can not remain real, since $\theta = 
-i\tilde{\theta} = -i \pi/4$ and $\tilde{\a} + \tilde{\b} = 2i c$, where
$c$ is real. Similar argument holds also for the two parameter solutions
with $G_-(r)$ as the non-extremality function.
Thus we can not get real time dependent solutions 
by Wick rotation in these cases and this is consistent with the 
observation made in
ref.\cite{br}.

\sect{Another class of solutions}

In this section we will discuss a different class of solutions of the equations
of motion than those discussed in section 2. Here also we relax the
supersymmetry condition and the function $A(r)$ and $B(r)$ appearing in the
metric will be taken to satisfy (2.8). However, we will see that the
solutions in this case can be reduced to supersymmetric solutions by a
coordinate transformation. We will recognize the solutions to be the
near horizon limits of the various BPS $p$-branes in $d=10$.

We have seen in section 2 that the equations of motion dictate that the
function $G(r)$ defined in (2.8) must satisfy eq.(2.16). If we do not
insist that $G(r)$ goes to unity asymptotically then the solution for
$G(r)$ can take the form,
\be
G(r) = \frac{{\hat \omega}^{2(q-1)}}{r^{2(q-1)}} = {\hat H}(r)^2
\ee
where ${\hat H}(r) = {\hat \omega}^{q-1}/r^{q-1}$ is a harmonic function
in the $(q+1)$-dimensional transverse space. Now comparing eqs.(2.13) and
(2.15) we find,
\be
\phi = \frac{a(d-2)}{q-1}B
\ee
Let us now make an ansatz for $B$ as
\be
B(r) = {\hat \alpha}\ln {\hat H}(r)
\ee
where ${\hat \alpha}$ is a parameter to be determined from the equations of
motion. Now substituting (4.2) and (4.3) into the equation for the function
$B$ in (2.13), we find that the equation simplifies to,
\be
{\hat \alpha} (q-1) {\hat \omega}^{2(q-1)} = \frac{b^2}{2(d-2)}
\frac{{\hat \omega}^{({\hat \alpha}\chi -2)(q-1)}}{r^{({\hat \alpha}\chi
-2)(q-1)}}
\ee
This equation can be solved if ${\hat \alpha} \chi = 2$ and the solution is
\be
{\hat \omega}^{q-1} = \sqrt{\frac{b^2\chi}{4(q-1)(d-2)}}
\ee
We note that this is exactly the form of ${\bar \omega}$ for the BPS D$p$
brane solutions we have defined earlier i.e., ${\hat \omega} = {\bar \omega}$.
It can be easily checked that the $R_{rr}$ equation (2.12) is automatically
satisfied with this solution. We thus find\footnote{We point out that the
corresponding time-dependent solutions given in \cite{br} has some 
typographical errors in eqs.(4.5) -- (4.7). We here correct them by comparing 
with eqs.(4.6) and (4.7) given below.},
\bea
e^{2B} &=& {\hat H}^{\frac{4}{\chi}}\nn
e^{2A} &=& {\hat H}^{-\frac{4(p+1)}{(q-1)\chi}+\frac{4}{q-1}}
\eea
The complete solution therefore takes the form,
\bea
ds^2 &=& {\hat H}^{-\frac{4(p+1)}{(q-1)\chi} + \frac{4}{q-1}}
\left(dr^2 + r^2
d\Omega_{d-p-2}^2\right) + {\hat H}^{\frac{4}{\chi}}\left(-dt^2 +
dx_1^2 + \cdots + dx_{p}^2\right)\nn
e^{2\phi} &=& {\hat H}^{\frac{4a(d-2)}{(q-1)\chi}}\nn
F_{[q]} &=& b\,\,{\rm Vol}(\Omega_{d-p-2})
\eea
and for the electric brane
\be
A_{[p+1]} = \sqrt{\frac{4(d-2)}{(q-1)\chi}} {\hat H} dt \wedge dx_1 \wedge
\ldots \wedge dx_p
\ee
One might be tempted to think that the solutions (4.7), (4.8) are
non-supersymmetric as the functions $A(r)$ and $B(r)$ appearing in the
metric do not satisfy the supersymmetry condition (2.7) rather they
satify (2.8) where $G(r) = \left(\frac{{\bar \omega}^{q-1}}{r^{q-1}}\right)^2$.
However, we show that by a coordinate transformation, the configurations
given in (4.7) and (4.8) can be cast into the form of the near-horizon limits
of various D$p$-branes confirming that the solutions are indeed supersymmetric.
We will mention also what happens to the condition (2.8) under this coordinate
transformation. Let us write the configurations (4.7) and (4.8) as,
\bea
ds^2 &=& \left(\frac{r}{{\bar \omega}}\right)^{\frac{4(p+1)}{\chi}}
\frac{{\bar \omega}^4}{r^4}\left(dr^2 + r^2
d\Omega_{d-p-2}^2\right) + \left(\frac{r}{{\bar \omega}}\right)
^{-\frac{4(q-1)}{\chi}}
\left(-dt^2 +
dx_1^2 + \cdots + dx_{p}^2\right)\nn
e^{2\phi} &=& \left(\frac{r}{{\bar \omega}}\right)^{-\frac{4a(d-2)}{\chi}}\nn
F_{[q]} &=& b\,\,{\rm Vol}(\Omega_{d-p-2}), \qquad
A_{[p+1]}\,\,\, =\,\,\, \sqrt{\frac{4(d-2)}{(q-1)\chi}} \frac{r^{-(q-1)}}
{{\bar \omega}^{-(q-1)}} dt \wedge dx_1 \wedge \ldots \wedge dx_p
\eea
Now let us make a coordinate transformation
\be
r = \frac{{\bar \omega}^2}{z}
\ee
Then we find
\be
\frac{{\bar \omega}^4}{r^4}\left(dr^2+r^2d\Omega_{d-p-2}^2\right)
= \left(dz^2 + z^2 d\Omega_{d-p-2}^2\right)
\ee
and we can rewrite (4.9) as,
\bea
ds^2 &=& {\bar H}(z)^{\frac{4(p+1)}{(q-1)\chi}}
\left(dz^2 + z^2
d\Omega_{d-p-2}^2\right) + {\bar H}(z)^{-\frac{4}{\chi}}
\left(-dt^2 + dx_1^2 + \cdots +
dx_{p}^2\right)\nn
e^{2\phi} &=& {\bar H}(z)^{-\frac{4a(d-2)}{(q-1)\chi}}\nn
F_{[q]} &=& b\,\,{\rm Vol}(\Omega_{d-p-2}), \qquad A_{[p+1]}\,\,\,=\,\,\,
\sqrt{\frac{4(d-2)}{(q-1)\chi}} {\bar H}(z)^{-1} dt \wedge \ldots \wedge
dx_p
\eea
where ${\bar H}(z) = \frac{{\bar \omega}^{q-1}}{z^{q-1}}$ is the $z \to 0$
limit of the harmonic function we defined in section 3. We recognize 
\cite{malda,imsy} (4.12) in
$d=10$ as the near-horizon limits of the BPS D$p$-branes and so the solutions
(4.12) are indeed supersymmetric. A similar solution can also be found for
the case of D-instanton. As before there is no $B(r)$ in the metric and
$A(r)$ would be given as,
\be
A(r) = \frac{1}{d-2}\ln G(r)
\ee
where
\be
G(r) = \frac{\hat{\omega}^{2(d-2)}}{r^{2(d-2)}} = \hat{H}^2(r)
\ee
Now using the ansatz,
\be
e^{\phi} = (\hat{H}(r))^{\hat {\nu}}
\ee
in eqs.(2.47) and (2.48) we obtain,
\be
\hat{\nu} = \frac{2}{a}, \qquad b = \frac{2}{a} (d-2) \hat{\omega}^{d-2}
\ee
The D-instanton solution therefore takes the form,
\bea
ds^2 &=& \frac{\hat{\omega}^4}{r^4}\left(dr^2 + r^2 
d\Omega_{d-1}^2\right)\nn 
e^{2\phi} &=& (\hat{H}(r))^{\frac{4}{a}}\nn
F_{[d-1]} &=& b {\rm Vol}(\Omega_{d-1}), \qquad A_{[0]} = \frac{2}{a}
\hat{H}(r) 
\eea
Again by making a coordinate transformation (4.10) we can rewrite the above
configuration as, 
\bea
ds^2 &=& \left(dz^2 + z^2 
d\Omega_{d-1}^2\right)\nn 
e^{2\phi} &=& (\bar{H}(z))^{-\frac{4}{a}}\nn
F_{[d-1]} &=& b {\rm Vol}(\Omega_{d-1}), \qquad A_{[0]} = \frac{2}{a}
\bar{H}(z)^{-1} 
\eea
where $\bar{H} = \bar{\omega}^{d-2}/z^{d-2}$ with $\bar{\omega} = 
\hat{\omega}$ and is the $z \to 0$ limit of the usual harmonic function. 
Eq.(4.18) is precisely the near horizon limit of the BPS 
D($-1$)-brane solution in $d=10$.
We can also obtain the corresponding time-dependent solutions obtained in
\cite{br} by the Wick rotation discussed earlier.

Let us now try to understand what happens to
the condition (2.8), i.e., $(p+1) B(r) + (q-1) A(r) = \ln G(r)$. Actually,
when we make a change of variable from $r \to z$, $\ln G(r)$ gets absorbed
into the function $A(r)$ i.e.,
\be
{\tilde A}(z) = - A(r) + \frac{1}{q-1} \ln G(r)
\ee
where $e^{{\tilde A}(z)}$ is the factor multiplying the transverse part of
the brane i.e., $(dz^2 + z^2 d\Omega_{d-p-2}^2)$. Thus in terms of
$z$-coordinate the relation (2.8) reduces to (2.7) i.e., supersymmetric
condition. In other words,
\be
(p+1) B(r) + (q-1) A(r) = \ln G(r) \quad \Rightarrow \quad (p+1)B(z)
+ (q-1) {\tilde A}(z) = 0
\ee
This happens only for this particular form of the harmonic function ${\hat H}
(r) = \frac{{\bar \omega}^{q-1}}{r^{q-1}}$ and $G(r) = {\hat H}(r)^2$. Thus
we clarify the reason why starting from the non-supersymmetric gauge condition
(2.8) we end up with supersymmetric solutions. Note that a similar solution
in \cite{br} for the time-dependent case were inappropriately called as the
non-BPS E$p$-branes in type II$^\ast$ string theories \cite{hull}. 
But by a similar
argument as presented here, those solutions should be the near horizon limits
of E$p$-branes in type II$^\ast$ string theories and they are supersymmetric.

\section{Conclusion}

In this paper we have constructed and discussed various aspects of the static,
non-supersymmetric $p$-brane (for $-1\leq p \leq 6$) solutions of 
type II supergravities in diverse dimensions. We have solved the equations
of motion of the relevant supergravity action by relaxing the supersymmetry
(or extremality) condition by introducing a non-extremality function. 
Equations of motion dictate the three specific forms of the non-extremality
function and hence the three specific classes of non-supersymmetric $p$-brane
solutions. We have explicitly constructed all three classes of solutions.
First two classes where the solutions are asymptotically flat are discussed
in section 2, whereas the last class where the solutions are not asymptotically
flat is discussed in section 4. We have also discussed the singularity 
structure and the region of validity of these solutions. Only one class of 
solutions discussed in section 2 were known before 
\cite{zz,im} in a different form
and we have clarified the relations of these solutions to those obtained
here in section 2. As $p=-1$ or D-instanton solution is different from the
other $p$-brane solutions, it is treated separately. Then we have discussed
how a subclass of these solutions (with $G_-(r)$ as the non-extremality 
function) can be interpreted as the interpolating solutions between
the $p$-brane--anti $p$-brane system and the usual BPS $p$-branes.
In order to obtain BPS $p$-branes the parameters of the non-supersymmetric
solutions were scaled in two different ways for $0\leq p \leq 6$ and a unique
way for $p=-1$. Then we also obtained BPS limits of the solutions with 
$G_+(r)$ as the non-extremality function. These give some unusual BPS brane
solutions. We have also shown how the real time-dependent solutions or
S$p$-brane solutions (including S($-1$)-brane) can be obtained from our
static solutions by a Wick rotation. We have seen that this happens only for
one class of solutions in section 2,
but for the other class Wick rotation does not give real time-dependent 
solutions. For the third class of solutions, we found that although apparently
the solutions are non-supersymmetric, however, by a coordinate transformation
we have seen that they are nothing but the near-horizon limits of various 
BPS $p$-branes
already known. We have given reasons why this happens for this particular
class of solutions. Finally, we point out that the solutions we constructed in
this paper have singularities. It would be interesting to understand 
the nature of the singularities and the ways, if at all possible, to 
resolve them.  
\vskip .5cm

\noindent
{\bf Acknowledgements}

    JXL would like to thank the Michigan Center for Theoretical Physics
 for hospitality and partial support during the final stage of this
 work. He also acknowledges support by grants from the Chinese Academy
 of Sciences and the grant from the NSF of China with Grant No: 90303002.

\end{document}